**Title**
One person's modus ponens…: Comment on "The Markov blanket trick: On the scope of the free energy principle and active inference" by Raja and colleagues (2021)

**Author**
Maxwell J. D. Ramstead

**Affiliations**
1. VERSES Research Lab, Los Angeles, California, 90016, USA
2. Wellcome Centre for Human Neuroimaging, University College London, London WC1N 3AR, UK

**Abstract**
In this comment on "The Markov blanket trick: On the scope of the free energy principle and active inference" by Raja and colleagues (2021) in *Physics of Life Reviews*, I argue that the argument presented by the authors is valid; however, I claim that the argument contains a flawed premise, which undermines their conclusions. In addition, I argue that work on the FEP that has appeared since the target paper was published underwrites a cogent response to the issues that are raised by Raja and colleagues.

I am very pleased to comment on "The Markov blanket trick: On the scope of the free energy principle and active inference" by Raja and colleagues (2021). The target paper poses interesting conceptual challenges to the free energy principle (FEP). Since it was preprinted in March 2021, it has been part of a productive and friendly critical dialectic. The target paper is noteworthy for its careful technical unpacking of one of the core versions of the FEP, formulated in terms of probability density dynamics over states. Here, I argue that the main argument in Raja et al. (2021) is valid, but that it contains a premise that we now know is flawed, which undermines their conclusions. In addition, I believe that work on the FEP that has appeared since the target paper was published underwrites a cogent response to the issues that it raises.

First, I agree with the main message of the paper, on the "Markov blanket trick." According to the authors, the FEP is at its core a modeling method, which leverages the fact that, if there is a Markov blanket in the nonequilibrium steady state density of a system, then it can be described as engaging in variational inference about the causes of its sensory perturbations. In other words, the FEP deploys a specific concept of "thing," defined as a cohesive locus of states enshrouded by a Markov blanket, which couples that thing to its embedding environment. When so described, an FEP-theoretic thing is a subset of a larger system endowed with a particular partition; where we can define a Markov blanket between two ("internal" and "external") subsets of the system, and where internal states track (or equivalently, in the present context, become the sufficient statistics of probabilistic beliefs about) external states.

Raja and colleagues' main argument against the applicability of the FEP has the form of a modus tollens (i.e., denying the consequent: $p \rightarrow q$ is equivalent to $\neg q \rightarrow \neg p$). Paraphrasing their argument, we have:
1. If a system is describable by the FEP ($p$), then it must have a Markov blanket ($q$)
2. Some interesting systems do not have Markov blankets ($\neg q$)
3. Therefore, those interesting systems must not be describable by the FEP ($\neg p$)

Their argument is valid. However, as the saying goes, "one person's modus ponens is another person's modus tollens." Recent work undermines the second premise of their modus tollens. Indeed, since the publication of the target paper in 2021, Friston and colleagues' sparse coupling conjecture for complex systems (first proposed in Friston et al., 2021) has been proven mathematically (Sakthivadivel, 2022a; see also Heins and Da Costa, 2022). We now know that, in the limit of increasing dimensionality, essentially all systems (both linear and nonlinear) will have Markov blankets, in the appropriate sense. That is, as both linear and nonlinear systems become increasingly high-dimensional, the probability of finding a Markov blanket between subsets approaches 1. Of note, in this work, the construct of Markov blanket has been weakened to "soft blankets," which are not entirely sealed off from the outside world. More technically, this work defines a "blanket index," which measures the distance between a perfectly blanketed system and a perfectly densely coupled system, such that we have a strict Markov blanket when the blanket index is zero. And we can show that this index tends to zero as systems increase in size. This formulation of Markov blankets, which allows for a separation of temporal scales and material turnover, may help address issues raised by the authors about meta-stable boundaries of dynamically changing systems, such as candle flames.

Raja and colleagues also claim that the FEP is ambiguous about whether it applies to the whole system (i.e., to the thing and its embedding environment) or if it restricts itself to describing things. Admittedly, the language used is sometimes ambiguous. But mathematically, there is no ambiguity. FEP-theoretic formulations formally distinguish between the system and the particle (i.e., particular states) that is part of the system—albeit, sometimes, less clearly than would be ideal. This issue has been addressed explicitly (e.g., Ramstead, Sakthivadivel et al., 2022): the generative models of the FEP are defined over the entire coupled brain-body-environment system. Defining a mechanics for the entire system in this way then allows us to describe the autonomous states of a particle as engaging in inference about external states. Thus, the FEP is a scientific model (implemented as a generative model of states or Lagrangian of paths), the content of which is precisely a mechanistic theory for the representational capacities of certain kinds of particles.

It is a truism in physics that to be a thing, physically speaking, entails separability (see Sakthivadivel 2022a for a formal argument to this effect, which appeals to statistical sampling). In the context of classical physics (Fields et al., 2022), separability means having a degree of conditional independence from the environment (i.e., having a Markov blanket). Contrary to what is claimed in the target article, then, the notion of thing defined under the FEP does follow from principled reflection on what it means to be a thing, physically.

Raja et al. raise an interesting point about the labels "internal" and "external." They claim that relational properties cannot always be cashed out using this dichotomy. And indeed, this terminology can be confusing. Mathematically, internal states are those states that encode the parameters of beliefs about other states, designated as external states. One might suggest that a better way of reading "internal" and "external" is "tracking" and "tracked." For instance, in active inference models of interoception, although the interoceptive states that are tracked are internal to the organism, they count as external states from the point of view of internal brain states. More generally, relational properties tracked by a particle are simply external states—and thus, do not pose a challenge to the formalism. In other words, relationships among external states are recapitulated in terms of relationships among internal states, which encode conditional dependencies among external states.

Raja and colleagues further claim that the FEP is not special, i.e., that it is just one among several alternative formalisms that might be used to model complex systems. Revealingly, they claim that the maximum entropy principle (for instance) could be used as an alternative to FEP. This choice is telling: it has been shown that FEP is dual to the constrained maximum entropy principle (CMEP), which is a cornerstone of contemporary physics (Sakthivadivel, 2022b; Ramstead, Sakthivadivel, et al., 2022). They are simply two complementary perspectives on the same phenomenon of self-organization: an FEP-theoretic model is a mechanical theory written to explain the dynamics of a system from the point of view of the "self" or particle; while the CMEP looks at the same phenomenon from the point of view of "organization," i.e., from the perspective of an external observer, in the environment or heat bath. An FEP-theoretic model of the dynamics of beliefs encoded by the states of a particle can thus always be converted to an equivalent CMEP-theoretic model of the dynamics of the states of the system. When we construct a maximum entropy model of the states of a system, we are also, ipso facto, constructing a model of the beliefs carried by the particle; and our FEP-theoretic model of a particle's belief updating is equivalent to an optimal model of the dynamics of a system, from the perspective of an outside observer. (Indeed, we can rederive the same kind of mode-matching and mode-tracking behavior that characterizes FEP-theoretic agents directly by applying maximum entropy inference to the beliefs parameterised by the states of the particle; see Ramstead, Sakthivadivel et al., 2022; Sakthivadivel, 2022b.)

This has a hugely significant implication: an FEP-theoretic model of a particle's beliefs about a system is a maximum entropy model of the states of the system. And a maximum entropy model of any system simply is the optimal model of that system, given our state of knowledge about that system (Jaynes 1957). In other words, the FEP-theoretic model of belief updating is the optimal model of systemic dynamics, given our current knowledge about the system. (I note that a closely related point was made earlier by Kiefer, 2020.)

Finally, a few remarks about formalism are in order. The argument on offer is correct, to a large extent, about the limitations of the density dynamics formulation of the FEP, which appeals to nonequilibrium steady states. But this represents only one of several ways in which the FEP has been applied (see Ramstead, Sakthivadivel, et al., 2022). Another point worth raising is that the FEP is not committed to ergodicity, in the generic statistical sense. In the density dynamics formulation, one can formulate a notion of "local ergodicity," as the kind of recurrence that characterizes the dynamics of systems endowed with attractors (Sakthivadivel 2022b). Thus, the FEP does not commit one to the idea that self-organizing systems need to explore all the states in their state space (which, with good reason, has been criticized, e.g., by Kauffman, 2019). Rather, the FEP implements the idea that systems (and the constituent particles) have some measurable characteristics. And indeed, arguably, the majority of biological processes subserve the maintenance of setpoints. The fact that these are dynamically modulated by allostatic processes—and undergo nonergodic drift—does not negate this fact; rather, it requires that one consider additional structure in the setting and maintenance of setpoints.

## Acknowledgments

I am thankful to two of the authors of the target paper, Vicente Raja and Anthony Chemero, for valuable discussions about their work. I am also thankful to Karl Friston, Mahault Albarracin, Alex Kiefer, and Dalton Sakthivadivel for discussions of this commentary.## References

Fields, C., Friston, K., Glazebrook, J. F., & Levin, M. (2022). A free energy principle for generic quantum systems. Progress in Biophysics and Molecular Biology.

Friston, K. (2019). A free energy principle for a particular physics. arXiv preprint arXiv:1906.10184.

Friston, K., Heins, C., Ueltzhöffer, K., Da Costa, L., & Parr, T. (2021). Stochastic chaos and markov blankets. Entropy, 23(9), 1220.